\documentclass[showpacs,preprintnumbers,amsmath,amssymb,twocolumn]{revtex4}

\usepackage{epsfig}

	     \usepackage{graphicx}

\usepackage{dcolumn}

\usepackage{bm}

\newcommand{\ea}{\end{eqnarray}}

\newcommand{\beq}{\begin{equation}}
\newcommand{\eeq}{\end{equation}}
\newcommand{\bea}{\begin{eqnarray}}
\newcommand{\eea}{\end{eqnarray}}

\newcommand{\gsim}{\lower.7ex\hbox{$
\;\stackrel{\textstyle>}{\sim}\;$}}
\newcommand{\lsim}{\lower.7ex\hbox{$
\;\stackrel{\textstyle<}{\sim}\;$}}

\def\cp{{\bf CP}}

\begin{document}

\title{Four-Quark Mesons in Non-leptonic $B$ Decays \\
--Could They Resolve Some Old Puzzles?}
\author{I.~Bigi}\email{ibigi@nd.edu}
\affiliation{Dept. of Physics, University of Notre Dame du Lac, Notre Dame, IN 46556, U.S.A.}
\author{L. Maiani}\email{luciano.maiani@roma1.infn.it}
\affiliation{Universit\`{a} di Roma `La Sapienza' and I.N.F.N., Roma, Italy}
\author{F. Piccinini}\email{fulvio.piccinini@pv.infn.it}
\affiliation{I.N.F.N. Sezione di Pavia and Dipartimento di Fisica Nucleare 
e Teorica, via A.~Bassi, 6, I-27100, Pavia, Italy}
\author{A.D. Polosa}\email{antonio.polosa@cern.ch}
\affiliation{Dip. di Fisica, Universit\`{a} di Bari and I.N.F.N., Bari, Italy}
\author{V. Riquer}\email{veronica.riquer@cern.ch}
\affiliation{I.N.F.N. Sezione di Roma}

\date{\today}

\begin{abstract}
We point out that non-leptonic $B$ decays driven by $b\to c \bar cs$ should provide 
a favourable environment for the production of hidden charm diquark-antidiquark 
bound states that have been suggested to explain the resonances with masses around 4 GeV recently observed by BaBar and BELLE. Studying their relative abundances in non-leptonic $B$ decays can teach us novel lessons about their structure and the strong interactions. Through their decay into 
$\psi$ they can provide a natural explanation of the excess of $B \to \psi X$ observed for 
$p_{\psi} < 1$ GeV. Other phenomenological consequences are mentioned as well. \newline
LNF-05/17(P), UND-HEP-05-BIG03, ROMA1-1417/2005, BA-TH/523/05
\end{abstract}

\maketitle
\section{Introduction}
Charm -- it seems -- is surprising us again. Introduced in 1970 to cure phenomenological deficiencies in the emerging Standard Model~\cite{GIM}, charm is now teaching us novel lessons about the strong interactions. The very recent discovery~\cite{Dsstar} of the $j_q = 1/2$ P wave charm-strange resonance $D_s^{**}$ with a mass significantly lower than expected has lead to a renaissance of charm spectroscopy. It was followed by the observation of some meson resonances with {\em no open} charm in $e^+e^-$ 
annihilation with masses around 4 GeV starting with $X(3872)$~\cite{BELLE3872} and so far ending with $Y(4260)$~\cite{BABAR4260}. 

Three different frameworks have been suggested to accommodate these states with their unusual 
characteristics: (i) $D-D^*$ molecules~\cite{molec}; (ii) $\bar ccg$ hybrids~\cite{PeneClose}; (iii) Diquark-antidiquark or four-quark states for short~\cite{MPPR}. 

The emerging proliferation of such states -- while surprising in scenarios (i) -- is quite natural in scenario (iii). Furthermore these states have been observed decaying into $\psi$ together with  light flavour hadrons, not necessarily as the dominant, yet as a significant channel. Since the $\psi$ represents a rather compact hadron, these decays are more natural in scenario (iii) than 
in the molecule picture, where $c$ and $\bar c$ are more separated spatially. 

Much more work both on the theoretical and the experimental side needs to be done to confirm these 
sightings, clarify the spectroscopy of the four-quark states and find additional ones. In this note we want to address two main points: (i) {\em Non}-leptonic $B$ decays should provide a favourable environment for the production of such states in marked contrast to semileptonic $B$ transitions. 
(ii) Such production will produce footprints in the final states, one of which might have been observed 
already, namely an excess of low-momentum $\psi$ in $B \to \psi +X$ over expectations.

This note will be organized as follows: after introducing the model of four-quark hadrons and their production in $B$ non-leptonic decays in Sect. \ref{fourquark}, we will analyze in Sect. \ref{FOOT} how decays of such four-quark mesons will  affect $B \to \psi +X$ and $B \to D_{(s)}/\bar D_{(s)}+X$, respectively, before concluding in Sect. \ref{SUM}. 

\section{Hidden Charm Four-Quark Hadrons, their Decays \& Non-leptonic $B$ Decays }
\label{fourquark}

The four-quark states we are referring to are diquark-antidiquark pairs in colour ${\bar 3}$ and $3$ 
configurations, respectively, bound together by colour forces. Thus they are quite distinct from 
$\bar D-D^*$ molecules held together by short range, colourless meson exchange forces. In particular our four-quark states are roughly of hadronic size. 

There is a full flavour $SU(3)$ nonet of such states with hidden charm, namely: 
\begin{itemize}
\item 
four with neither open nor hidden strangeness: $X_{q\bar q} = (cq)(\bar c \bar q)$, $q=u,d$; 
\item 
four with open strangeness ($S=\pm 1$): $X_{s\bar q} = (cs)(\bar c \bar q)$, 
$X_{q\bar s} = (cq)(\bar c \bar s)$, 
$q=u,d$;
\item 
one with hidden strangeness: $X_{s\bar s} = (cs)(\bar c \bar s)$. 

\end{itemize}
Furthermore these combinations can come as S waves and their orbital excitations. 

The S wave states have been analyzed in Ref.~\cite{MPPR} under the assumption that inter-diquark forces are approximately spin independent due to the large value of the charmed quark mass. In this case, diquark and antidiquark total spin can take both $S=0, 1$, and a large multiplet results. 
S wave states have positive parity and  $J^{PC}=0^{++}(2), 1^{++}, 1^{+-}(2), 2^{++}$ (in parenthesis, the multiplicity of the given $J^{PC}$). 
For the recently observed $X(3872)$ and $X(3940)$ resonances the S wave, $X_{q\bar q}$ assignment has been suggested with $J^{PC}=1^{++}, 2^{++}$, respectively. The presence of $u$ and $d$ quark makes it possible to have large violations of G parity in the wave functions, accounting for the {\em simultaneous} presence of  the decay modes ~\cite{BELLE2}: 
\begin{equation}\label{psidec}
X(3872)\to \psi+ 2\pi , \psi+ 3 \pi
\end{equation}

Unnatural spin-parity  forbids  $D{\bar D}$ decays of the $X(3872)$ and justifies a sizable decay  into charmonium+ light hadrons. The quark spin combination in the $X(3872)$ is such that the $c{\bar c}$ pair has total spin $S_{c{\bar c}}=1$. Conservation of the heavy quark spin implies that the decay into 
$\psi + V$ is allowed while $\eta_c + P$ is forbidden~\cite{MPPR,Voloshin}  (V=Vector  and P=Pseudoscalar nonets). 
The competitive decay channel is:
\begin{equation}\label{D0dec}
X(3872)\to D^0 ({\bar D}^0)  +{\bar D}^{*0}(D^{*0})\to D^{0}+{\bar D}^{0}+ \pi^0
\end{equation}
This channel is estimated in~\cite{M_etalnew} to have a rate about equal to 
that for $\psi+\rho$. Within theoretical uncertainties the rate of (\ref{D0dec}) could be 
perhaps three to five times larger, so we can estimate:
\beq
B(X(3872)\to\psi+{\rm all})>0.3.
\eeq

There are two classes of {\em strange} counterparts: (i) $X_{u(d), \bar s}$ can arise in  
$B^+ \to X_{u(d),\bar s} \pi + ... \to (\psi K \pi)\pi +...$, where the ellipses denote additional pions. 
(ii) $X_{s,\bar u (\bar d)}$ lead to the unusual final states  
$B^+ \to X_{s,\bar u (\bar d)}  K  K +... \to (\psi \bar K \pi ) K  K +...$.   

Small widths and sizable decay fractions into charmonium are also expected for the lightest scalar state, which is predicted to be below threshold for $D{\bar D}$ decay, and for the other two  axial mesons, because of unnatural spin-parity. However, heavy particle spin conservation now permits both decays with $\psi$ and $\eta_c$:
\begin{eqnarray}\label{0,1dec}
0^{++}\to\psi+V, \eta_c+P, \nonumber  \\ 
1^{+-}\to\psi+P, \eta_c+V. 
\label{ETAC}
\end{eqnarray}

In conclusion, the S wave diquark-antidiquark multiplet contains in all four full nonets whose members have large branching fractions for decays into $\psi$~+ light hadrons, of order $0.5\cdot B(X(3872)\to\psi+{\rm all})$.

The recently observed $Y(4260)$~\cite{BABAR4260} was assigned to a  P wave $X_{s\bar s}$ with diquark and antidiquark spin $S=0$ in Ref.~\cite{MPPR2}. It should have a dominant decay into $D_s{\bar D}_s$ and therefore a small branching fraction into $\psi$, estimated around $0.05$ in Ref.~\cite{MPPR2}.

Non-leptonic $B$ decays should provide a dynamical environment relatively favourable to the production of four-quark states, if they are kinematically accessible. More specifically 
the $b \to c \bar c s$ transition, which drives about 25 - 30 \% of all 
$B$ decays (as can be inferred from the charm content of the final states in $B$ decays, which is  measured to be around $1.3$ \cite{NCCOUNT}) leads to three 
(anti)quarks {\em locally}, as far as hadronic distances are concerned, where $c \bar c$ in general do not form a colour singlet. 

\begin{figure}[ht]
\begin{center}
\epsfig{
height=4truecm, width=8truecm,
        figure=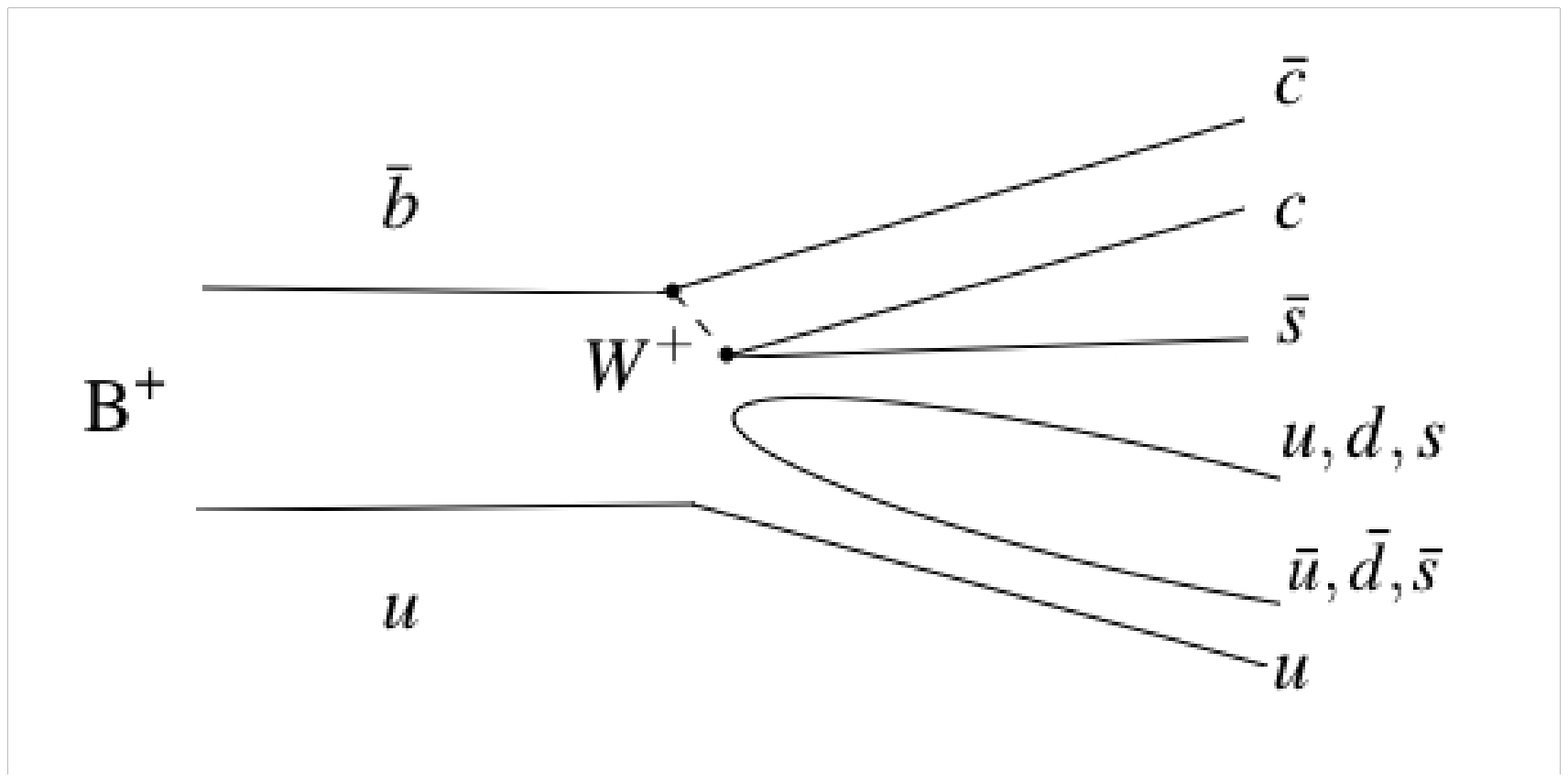}
\caption{Quark diagram for $B^+ \to K+X$, with $X=(c\bar c q {\bar q}').$\label{Bnonlept} \footnotesize 
}
\end{center}
\end{figure}

This should enhance considerably the probability that hadronization will lead to the emergence of four-quark states with hidden charm, for the addition of  a single quark or antiquark from hadronic fragmentation suffices, see Fig.~1. In semileptonic $B$ decays on the other hand those four-quark states should be practically absent.

To translate this qualitative expectation into a reliable numerical prediction is beyond our present capabilities. Measuring the relative abundances of the four-quark resonances in $B$ would provide 
us with novel information on the structure of these resonances and QCD's inner workings. 
For our present study we will use estimates that, however, may not be unreasonable. 
The $c$ and $\bar c$ are produced with small relative momenta. One can envision two 
different scenarios: (i) One assumes the $c \bar c$ hadronizes into four-quark states by picking 
up both a quark and an antiquark from fragmentation. The production of $X(3872)$ identified with 
an $1^{++}$ $X_{q\bar q}$ then provides a realistic yardstick. (ii) It might be tempting to 
argue that one starts primarily from a $(c\bar s)\bar c$ combination which could lead to 
the emergence of $X_{qs}$ states as the leading contribution. In that case the observed yield 
of $X(3872)$ provides an {\em under}estimate for the production of all four-quark resonances. 

Amplitudes for the general decay $B^+\to P+X$ can be derived from the diagram of Fig.~1. There being  in the final state two light quarks and two light antiquarks, the $P$ meson can arise from a $q{\bar q}$ pair in four possible ways. In the exact $SU(3)_{\rm flavor}$ limit, each of these ways correspond to one invariant coupling of the $\Delta I=0,~ \Delta S=+1,~\Delta C=0$ weak Hamiltonian and spectator $u$ (${\bar 3}$ and $3$, respectively) to the two final nonets. Explicitly, this leads to the following parametrization of decay amplitudes: 
\begin{eqnarray}\label{Bplusnld}
&&{\it M}(B^+\to P+X)= \nonumber \\
&&A[K^+ X_{u{\bar u}}+K^0\ X_{u{\bar d}} +( \sqrt{\frac{1}{3}}~\eta_0-\sqrt{\frac{2}{3}}~\eta_8 )X_{u{\bar s}}]\nonumber\\
&&+B~K^+(X_{u{\bar u}} + X_{d{\bar d}}+X_{s{\bar s}})  \nonumber\\
&&+ C~[\pi^+X_{d{\bar s}} + K^+~X_{s{\bar s}} + (\pi^0\frac{1}{\sqrt{2}}+\eta_8\frac{1}{\sqrt{6}} + \eta_0\frac{1}{\sqrt{3}})~X_{u{\bar s}}] \nonumber\\
&&+ D~\sqrt{3} ~\eta_0~X_{u{\bar s}}
\end{eqnarray}
where A, B, C, D are a priori unknown constants. The total decay rate then follows:
\begin{eqnarray}
\Sigma_{i,j} \Gamma (B^+\to P_i+X_j)&= &3 (A^2 + B^2 + C^2+ D^2)\nonumber\\
&+ &2(A+C)(B + D)
\label{Bplusrate}
\end{eqnarray}
Following Ref.~\cite{MPPR}, we assume that $X_{u{\bar u}}$ and $X_{d{\bar d}}$ are approximate mass eigenstates and we normalize rates to $\Gamma_{u{\bar u}}=\Gamma(B^+\to K^+ +X_{u{\bar u}})$, which is assumed to dominate $B^+$ decay. This is in agreement with the approximate observation of a single narrow structure in this decay~\cite{BELLE3872} and with the recent BaBar hint of a positive mass difference for the X-particle in $B^0$ w.r.t. the one in $B^+$ decays\cite{BaBar2}, $M(X_{B^0})-M(X_{B^+})=(2.7\pm1.3\pm0.2) $~MeV. 

It is simple to see then that:
\begin{eqnarray}
&&\Sigma_{i,j} \Gamma (B^+\to P_i+X_j)>\frac{15}{8}~\Gamma_ {u{\bar u}}
\label{liminf}
\end{eqnarray}

The minimum value is obtained for $A=B=1/2$, $C=D=-1/8$. Interestingly, this solution corresponds to one of the two decay patterns envisaged in Ref.~\cite{MPPR}, where $X_{u{\bar u}}$ dominates the decay $B^+\to K^+ X(3872)$:
\begin{eqnarray}\label{ratplus}
&&\Gamma(B^+\to K^+ +X_{u{\bar u}}): \Gamma(B^+\to K^+ +X_{d{\bar d}})\nonumber\\
&&:\Gamma(B^+\to K_S +X_{u{\bar d}})=1:1/4:1/8
\end{eqnarray}
In terms of the diagram of Fig.~1, the dominance of $u\bar u$ decay indicates that the dominant amplitude corresponds to $K^+$ formation by combining the $\bar s$ with the $u$-quark from the sea rather than with the spectator quark.

The analogous formula to Eq.~(\ref{Bplusnld}) for $B^0$ decay is obtained by exchanging $u$ and $d$ flavors: 
\begin{eqnarray}\label{Bzeronld}
&&{\it M}(B^0\to P+X)= \nonumber \\
&&A[K^0 X_{d{\bar d}}+K^+\ X_{d{\bar u}} +( \sqrt{\frac{1}{3}}~\eta_0-\sqrt{\frac{2}{3}}~\eta_8 )X_{d{\bar s}}]\nonumber\\
&&+B~K^0(X_{u{\bar u}}+ X_{d{\bar d}}+X_{s{\bar s}}) \nonumber\\
&&+ C~[\pi^-X_{u{\bar s}} + K^0~X_{s{\bar s}} + (-\pi^0\frac{1}{\sqrt{2}}+\nonumber\\
&&+\eta_8\frac{1}{\sqrt{6}} + \eta_0\frac{1}{\sqrt{3}})~X_{d{\bar s}}] + D~\sqrt{3} ~\eta_0~X_{d{\bar s}}
\end{eqnarray}

In correspondence to the previous solution, one has  that $X_{d{\bar d}}$ dominates $B^0\to K_S+X$ decay:
\begin{eqnarray}\label{ratzero}
&&\Gamma(B^0\to K_S +X_{u{\bar u}}): \Gamma(B^0\to K_S +X_{d{\bar d}})\nonumber\\
&&:\Gamma(B^0\to K^+ +X_{d{\bar u}})=1/4:\;1:\;1/2
\end{eqnarray}

The overall pattern is consistent with the fact that the exotic partners: $X^{\pm}=X_{u{\bar d}}, X_{d{\bar u}}$ have not been yet observed in B decays.

\section{Possible Footprints in Non-leptonic $B$ Decays} 
\label{FOOT}
Even if non-leptonic $B$ decays are a relatively rich source of hidden-charm four-quark states, it represents a formidable task to identify them directly and separately, in particular since one expects a host of them  being kinematically accessible. As an intermediate step one can search for more inclusive `footprints'  four-quark states can leave behind in the final states of $B$ decays. We consider two classes, namely the production of hidden and open charm hadrons, $\psi/\eta_c$ and $D_{(s)}/\bar D_{(s)}$. 

\subsection{$B \to [\bar cc] +X$}

The inclusive $\psi$ spectrum in $B$ decays has been studied very carefully for two main reasons: 
Channels like $B_d \to \psi K_S$, $\psi K_L$ etc. had been predicted to exhibit large \cp~asymmetries (correctly, as it turned out); the form of the $\psi$ spectrum has been predicted using 
NRQCD~\cite{NRQCD}. It was found that the NRQCD predictions agree quite well with the data over most of the range. However for `slow' $\psi$, i.e with momenta below 1 GeV, that data show a marked excess over expectations corresponding to a branching ratio of about 
$6 \times 10^{-4}$ \cite{EXCESS}. 

No explanation has been established yet for this excess. One 
suggestion was to invoke a $\sim 1\%$ `intrinsic charm' 
(IC) component in $B$ mesons, which would produce final states of the form $\psi DX$
\cite{ICEXPL}. Those have been searched for, yet not found. The upper bound is such that the IC option can safely be ruled out as a source for the excess of the soft $\psi$ \cite{BABARIC}. 

A very different proposal was made in Ref.~\cite{DYDEK}, where the excess of soft $\psi$ was 
attributed to the production of a hybrid meson carrying strangeness with a branching ratio 
$\sim 10^{-4}$: 
\beq 
B \to \psi K_{{\rm hybrid}} 
\eeq

Instead we propose that these soft $\psi$ represent footprints of the production of hidden charm four-quark states in $b\to c \bar c q$ decays and their subsequent decay into $\psi$ plus light-flavour hadrons. While the absolute numbers we can give are very uncertain, we believe they are in an a priori 
reasonable range and point to a very nontrivial phenomenology. 

We start from the experimental value:
\begin{eqnarray}\label{fract1}
&&B(B^+\to K^++X(3872))\times B(X(3872)\to \psi  +\pi^+ \pi^-)\nonumber\\
&&\simeq1\cdot 10^{-5},
\end{eqnarray}
\noindent The near equality of the braching ratios of the two modes (\ref{psidec}) implies $B(X(3872)\to\psi +{\rm light\;hadrons})\simeq 2 B(X(3872)\to \psi  +\pi^+ \pi^-)$~\cite{MPPR}, so that:
\begin{eqnarray}\label{fract1}
B(B^+\to K^++X) 
B(X\to \psi+{\rm \;l.\;h.})\simeq 2\cdot 10^{-5},
\end{eqnarray}
where (l.h.=light hadrons).
Assuming that all particles in the nonet of X(3872) have similar branching ratio for decays into $\psi$, one finds from Eq.~(\ref{liminf}) 
\begin{eqnarray}\label{fractX}
&&\Sigma_{i,j}B(B^+\to P_i+X_j) B(X_j\to \psi + {\rm l.h.}) > \nonumber \\
&&>1.9B(B^+\to K^+ +X) B(X\to \psi +{\rm l.\;h.})\\ \nonumber
&&\simeq 3.8\cdot 10^{-5}
\end{eqnarray}

Including the scalar and the axial multiplets $0^{++}, 1^{+-}$, under the assumption that they are produced in $B^+\to K^+ +...$ decays with similar rates as $X_{u{\bar u}}$ and that they have branching 
ratios in $\psi$ of the order of 50\% of the $X(3872)$ branching ratio, we may get an additional factor of 2.5, thus obtaining:
\begin{eqnarray} 
\label{fractK}
&&\Sigma_X [B(B^+ \to P +X_{S{\rm wave}})B(X_{S{\rm wave}}\to \psi +{\rm l.\;h.})]\nonumber\\
&&> 0.94 \times 10^{-4}
\end{eqnarray}

The estimate in Eq.~(\ref{fractK}) does not correspond yet to the overall inclusive B decay into a $\psi$ via four-quark states.  For $J=1$ four-quark states it is quite possible or even likely that decay modes such as:
\begin{eqnarray}
&&B\to X(J^{P,C}=1^{+,\pm})+V\;{\rm or}\;A
\end{eqnarray}
dominate, since they can occur in S-wave unlike those involving pseudoscalar mesons, which 
occur in P-wave. Such a case occurs in $B\to \psi + (K+n\pi)$, where $K$, $K^*$(892) and $K_1$(1270) modes are in the ratios~\cite{PDG}:
\begin{eqnarray}\label{ratpsi}
&&B(B\to \psi + K):\;B(B\to \psi + K^*):\;B(B\to \psi + K_1)\nonumber\\
&&=1:\;1.4:\;1.8
\end{eqnarray}
Accounting for decay modes with vector and axial vector light resonances can 
thus introduce an additional factor of four into the estimate in Eq.~(\ref{fractK}), which leads us to the 
{\it educated guess}:
\begin{eqnarray}\label{guess}
&&\Sigma_{X,M} [B(B^+ \to M +X_{S{\rm wave}})B(X_{S{\rm wave}}\to \psi +{\rm l.\;h.})] \nonumber\\
&&> 3.8\cdot 10^{-4}
\end{eqnarray}

In conclusion, we propose the following `generic' decay chain:
\begin{eqnarray}
\label{CHAIN}
&&B \to X(3872)K_i \to (\psi \pi^+\pi^-) K_i  \nonumber\\
&&{\rm with} \; \; K_i = K, K^*(892), K_1(1270)
\end{eqnarray}
In Fig.~2 we show the resulting spectrum of $\psi$ with momenta expressed in the 
$\Upsilon (4S)$ frame for the three cases $K$, $K^*(892)$ and $K_1(1270)$ 
separately as well as added up with the ratios given in Eq.~(\ref{ratpsi}). 

\begin{figure}[ht]
\begin{center}
\epsfig{
height=4truecm, width=8truecm,
        figure=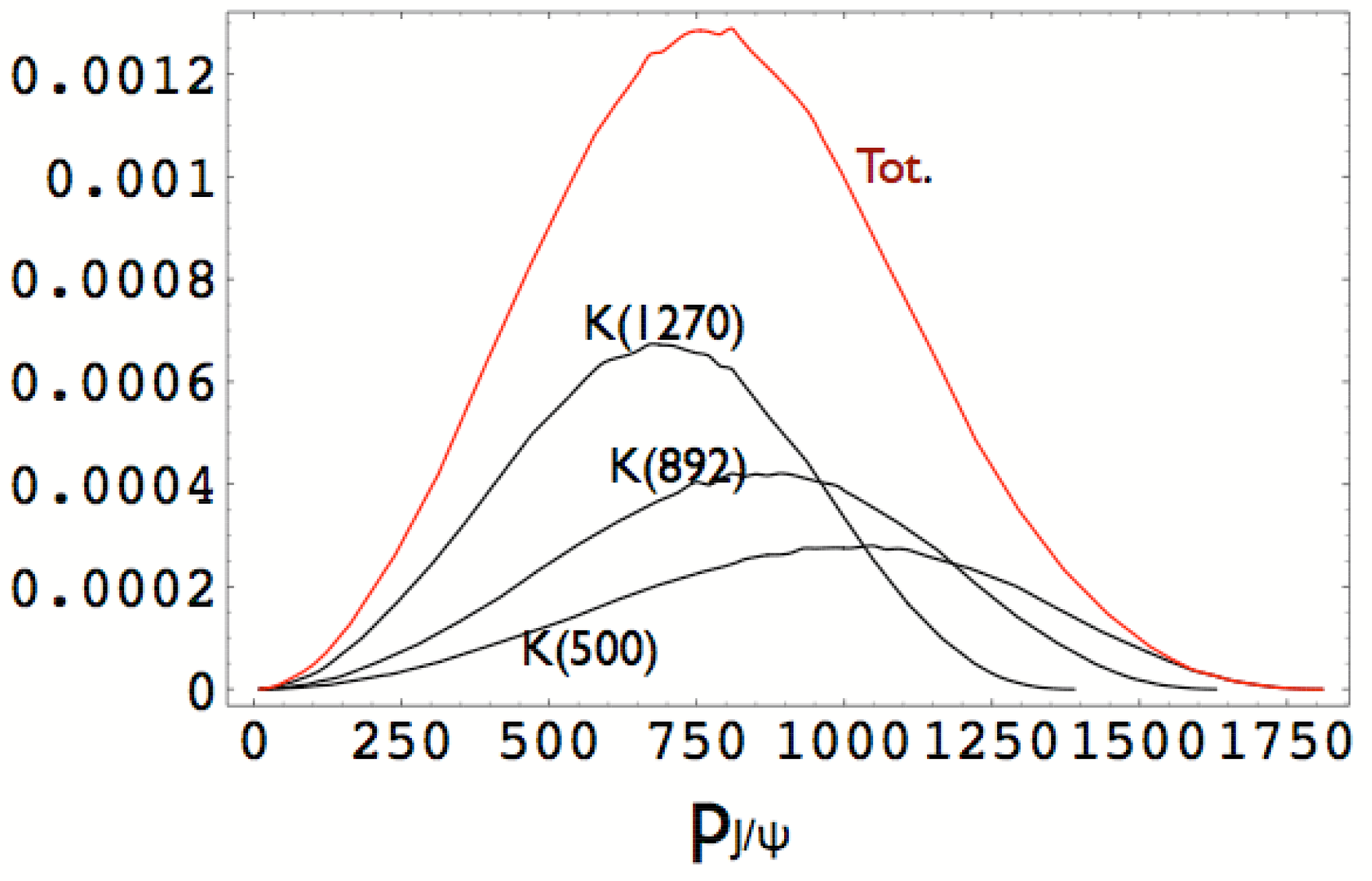}
\caption{$\psi$ momentum distribution in the rest frame of $\Upsilon$(4S) for the decay
$B\to X(3872)+K_i\to \psi \pi^+\pi^- +K_i;\; K_i=K(500),\;K(892),\;K(1270)$. The curve labeled $Tot.$ correspond to the sum of the three spectra weighted with coefficients as in Eq.~(\ref{ratpsi}) of text.}
\end{center}
\end{figure}

We have called this decay chain `generic', since strictly speaking it is typical for 
$X_{u\bar u, d\bar d,u\bar d,d\bar u}$ production only. For $X_{u(d),\bar s}$ one has, as 
already mentioned, $B^+ \to X_{u(d)\bar s} \pi +... \to (\psi K\pi)\pi +...$. Yet due to the 
smallness of the pion mass one can expect the $X_{u(d),\bar s}$ to recoil against at least 
two or more pions, i.e. the $\rho$ and/or $a_1$. This leads to a $\psi$ spectrum very similar to that 
from $B \to X(3872)K^*/K_1$. The intriguing $X_{u(d),\bar s}$ production with its 
unusual final state 
$B^+ \to X_{s,\bar u (\bar d)} K K +... \to (\psi \bar K \pi) K  K +...$ will yield  an even 
softer   $\psi$ spectrum. 

The estimated probability to produce a $\psi$ via four-quark mesons is of the same order as the observed excess of low-momentum $\psi$s ~\cite{BABARIC}. Such a large value was obtained from the (relatively safe) value for the $X(3872)$ by a multiplication factor of about 20. While this factor 'per se' is rather uncertain, its order of magnitude should be correct in that it simply reflects the large multiplicity of states predicted by the four-quark model to appear in B decay. These decays produce rather low 
momentum $\psi$, since the latter are a secondary decay product. If the four-quark states are typically 
produced in conjunction with a vector or axialvector light flavour hadron rather than a pseudoscalar 
one -- due to their $J^{PC}$ quantum numbers or due to other dynamical effects -- then the resulting 
$\psi$ production will peak {\em below} 1 GeV. 

As stated in Eq.~(\ref{ETAC}) the decays of scalar and axial four-quark states yield also 
$\eta_c$, one expects some excess also of soft $\eta_c$ in nonleptonic $B$ decays.



\subsection{$B \to D_{(s)} +X, \bar D_{(s)} +X$}
\label{DSDS}

While the $X(3872)$ is above $D\bar D$ threshold, its quantum numbers $1^{++}$ do not allow 
$X(3872) \to D \bar D$. Yet some of the other four-quark states can decay into a pair of open charm 
mesons, in particular $Y(4260) = [(cs)(\bar c \bar s)] \to D_s^+D_s^-$ should be dominant. Some of 
the other suggested four-quark resonances will have sizable decay rates into $D$ mesons as well. 

$B$ decays to conventional hadrons will produce charm mesons directly and frequently. In particular 
$\bar B_q = (b\bar q) \to D^+_s X$ as well as $\bar B_q = (b\bar q) \to D^-_s X$ are possible; 
the former is 
due to a $b \to c "W^-"$ transition with subsequent `popping' of an $\bar s  s$ pair, and the latter 
due to $b \to (\bar cs) c$. Yet the spectra in both cases are quite different and 
rather hard. They have been studied in a detailed way, in particular in the context of `charm 
counting', i.e. determining how many charm hadrons are produced in $B$ decays. 

The reaction $B \to X_{s\bar s} +(\bar K +\pi's) \to D_s^{+(*)}D_s^{-(*)} + (\bar K +\pi's)$ will produce 
a pair of $D_s$ mesons with very soft and {\em symmetric} spectra. Studies with even larger samples 
of $B$ mesons might reveal their presence~\cite{COUDERC}.

\section{Conclusions}
\label{SUM}

While the notion of multi-quark states has been around for a very long time going back to the early 
days of the quark model, heavy flavour dynamics have offered novel perspectives onto this notion, 
in their composition as well as in their production environment. Hidden-charm 
diquark-antidiquark resonances represent a good example for such perspectives: 
heavy quark symmetry employed for the charm quarks provides us with very useful theoretical tools, 
and $B$ decays constitute an intriguing production environment. 

Four-quark states should be practically absent from {\em semi}-leptonic $B$ decays, whereas 
{\em non}-leptonic transitions can be expected to provide a relatively fertile ground for their 
production. Establishing their presence in nonleptonic $B$ decays and measuring their 
relative abundances there will teach us important and novel lessons on the inner workings of QCD. 

Furthermore the production and decay of hidden-charm four-quark states would leave footprints in the 
final states of nonleptonic $B$ deccays. In particular it could resolve a long-standing puzzle, namely the excess of $\psi$ over expectations observed below momenta of 1 GeV. That excess has never been explained  successfully with conventional physics. It has 
been blamed on intrinsic charm; yet that explanation has been ruled out.

We cannot provide a numerically rigorous estimate for our explanation of the soft $\psi$ excess. 
However we consider the numbers we have presented above as rather reasonable and natural. 
Our ansatz reproduces not only the number of the excess $\psi$, but also -- and we consider that 
a very important point -- their momenta being 
mainly below 1 GeV in a natural way. 

To obtain these results, we had to call on several  core features of the diquark-antidiqurk picture: the existence of and contribution from several multiplets of such resonances; 
most of them having at least sizable branching ratios into $\psi$; their quantum numbers 
suggesting that these resonances are produced in nonleptonic $B$ decays mainly in conjunction 
with vector or axialvector light-flavour hadrons rather than pseudoscalar ones. While four-quark scenarios `fit the bill' in this sense, we do not see how $D-D^*$ molecule scenarios could shed any light 
on the problem of the excess $\psi$. 

We do not claim to have provided a compelling solution. Our proposal is meant as an invitation for further scrutiny in a new direction. We are aware of another possible solution to the slow 
$\psi$ puzzle \cite{DYDEK}. Ours is intimately connected with attempts to provide a 
consistent home to a number of newly discovered or at least observed hadronic resonances with 
masses around 4 GeV and having sizable decays into $\psi$. We have also identified 
further potential footprints that might surface in the future.


{\sl Acknowledgments}.
We thank R.~Faccini for interesting discussions. 
We would like to gratefully acknowledge the hospitality extended to some of us by the Frascati Spring Institute, Italy, (I. B., F.P., A.D.P.), and by the LPTENS, Ecole Normale Superieure, Paris, (L.M. and V.R). 
This work was supported by the NSF under grant PHY03-55098 and partially supported by Ministero dell' Istruzione, Universit\`{a} e Ricerca and Istituto Nazionale di Fisica Nucleare.

\end{document}